\documentclass[aps,twocolumn]{revtex4}
\usepackage[dvips]{graphicx}
\usepackage{amsmath, amssymb, bm}

\newcommand{\cp}{$CP^1$}
\newcommand{\n}{{\bf n}}
\newcommand{\z}{{\bf z}}
\newcommand{\bj}{{\bf j}}
\newcommand{\bl}{{\bf l}}
\newcommand{\A}{\mathcal {A}}
\newcommand{\ra}{\rangle} % bra
\newcommand{\la}{\langle} % ket

\begin{document}
\title{Emergent Photons and New Transitions in the O(3) Sigma Model
with Hedgehog Suppression.}
\author{Olexei I. Motrunich}
\author{Ashvin Vishwanath}

\affiliation{Department of Physics, Massachusetts Institute of
Technology, Cambridge MA 02139.}

\date{\today}
 
\begin{abstract}
{We study the effect of hedgehog suppression in the O(3) sigma model in
D=2+1. We show via Monte Carlo simulations that the sigma model can be
disordered while effectively forbidding these point topological
defects. The resulting paramagnetic state has gauge charged matter with
half-integer spin (spinons) and also an emergent gauge field
(photons), whose existence is explicitly demonstrated. Hence, this is
an explicit realization of fractionalization in a model with global
SU(2) symmetry. The zero temperature ordering transition from this
phase is found to be continuous but distinct from the regular
Heisenberg ordering transition. We propose that these phases and this
phase transition are captured by the {\it noncompact} $CP^1$ model,
which contains a pair of bosonic fields coupled to a
noncompact $U(1)$ gauge field. Direct simulation of the transition in
this model yields critical exponents that support this claim. The
easy-plane limit of this model also displays a continuous zero
temperature ordering transition, which has the remarkable property of
being self-dual. The presence of emergent gauge charge and hence
Coulomb interactions is evidenced by the presence of a finite
temperature Kosterlitz-Thouless transition associated with the thermal
ionization of the gauge charged spinons. Generalization to higher
dimensions and the effects of nonzero hedgehog fugacity are
discussed.}
\end{abstract}

\maketitle
\section{Introduction}
Since the initial proposal of fractionalization (phases where the
elementary excitations are fractions of the electron) as the
underlying explanation for the unusual properties of the cuprate
superconductors \cite{Anderson}, there has been much progress in the
theoretical understanding of these phases. Whether such fractionalized
phases in the absence of magnetic fields in spatial dimension greater
than one are actually realized in any experimental system is a matter
of current debate. Fractionalized phases can exhibit properties that
are strikingly different from conventional phases of matter, hence
they are attractive candidates for modeling strongly correlated
systems that exhibit anomalous behaviour. However, unambiguous
experimental evidence for the presence of such phases in any
experimental system is still lacking. In part this may be because the
correlations in such phases are subtle, and hence definitive
experimental signatures are hard to devise. This provides a strong
motivation to seek a deeper understanding of these
phases. Furthermore, fractionalized states have been proposed as a
means to build quantum bits that are inherently robust against
decoherence \cite{Kitaev}.

An important theoretical development has been the discovery of a
number of microscopic models \cite{kitaev, sondhi, balents, ioffe,
wengb, motrunich, wen, hermelle, huse} that can be shown to exhibit this
exotic physics. Although these microscopic models are defined in terms
of bosons (or spins) on a lattice with short ranged interactions,
fractional excitations that are charged under an emergent gauge field,
as well as excitations of this emergent gauge field are obtained on
solving these models. However till date, there have been no
microscopic models available with the full SU(2) spin rotation
symmetry. Indeed it is important to verify that the additional
constraints imposed by spin rotation symmetry do not exclude the
possibility of fractionalization. Here we will describe a model that
possesses the full spin rotation symmetry, but can be explicitly
shown to exhibit fractionalization and possess an emergent gauge field
in the deconfined phase. Moreover this model is found to have a
quantum critical point with full spin rotation symmetry, but which
is distinct from the Heisenberg transition. 
The properties of the fractionalized phase and the transition,
as well as various deformations on the model, will be studied in 
detail in this paper. 

Most of the earlier work that constructed models exhibiting
fractionalization engineered the energetics so as to select a low
energy manifold. Constraining states to lie in this manifold
introduces the gauge fields, which then need to be in the deconfined
phase for fractionalization to occur. Here we will rely on a
different route to fractionalization, which may be described as
fractionalization from defect suppression. Indeed this approach is
closer in spirit to \cite{NodalLiquids}, where the $Z_2$
fractionalized state was regarded as a quantum disordered
superconductor, obtained by proliferating even winding-number vortices while
suppressing vortices of odd winding number. 

Here we will mainly be concerned with the O(3) sigma model in
D=2+1. This model consists of O(3) quantum rotors represented by
unit three-vectors (`spins'), defined on the sites of a two dimensional
spatial lattice. Neighbouring rotors are coupled via a ferromagnetic
interaction. By the usual Quantum to Classical Statistical Physics
mapping, the ground state properties of this model can be conveniently
mapped onto the physics of the Heisenberg model at finite temperature
in three dimensions. Clearly, there exist point topological defects in
the three dimensional Heisenberg model that carry an integer
topological charge,
which simply correspond to hedgehog configurations of the spins. In
terms of the quantum model, these are events in space-time
(instantons) which change the skyrmion number of the system. We now ask
the question:  Is it possible to disorder the three dimensional
Heisenberg model in the effective absence of the hedgehog defects?
This has been a long standing issue, discussed in several works, for
example \cite{Dasgupta,KamalMurthy}, but had not been conclusively
settled. Here we will present fresh results from Monte Carlo
simulations and arguments that convincingly demonstrate that the
answer to this question is yes.  

Having established this, we will ask the question, what is the nature
of this hedgehog-free paramagnetic phase $P^*$? Although the spin-spin
correlations are short ranged, the absence of hedgehog fluctuations
lead to a `hidden order' in this phase. Indeed, it will be proposed
that the physics of the hedgehog-free model is captured by what we
will call the {\it non-compact} \cp (NC\cp) \cite{nccp1} model in
D=2+1. This model consists of a doublet of bosonic fields (`spinons')
that transforms as a spinor under spin rotations, coupled to a
non-compact U(1) gauge field (`photon'). In this representation, the
hedgehogs correspond to the monopoles of the U(1) gauge field, and
eliminating hedgehogs leads to the noncompactness of the gauge
field. The NC\cp model has two obvious phases, one where the spinons
are condensed which is the ferromagnetic phase, and the other where
the spinons are gapped, which corresponds to an exotic paramagnet, with a gapless photon excitation. To verify that the 
paramagnetic phase $P^*$ obtained by the hedgehog-free disordering of the O(3) sigma
model is indeed the same phase as the paramagnet in the NC\cp model, we use the Monte Carlo method to measure the
correlations of the spin chirality, which is roughly
$\n_1\cdot(\n_2\times \n_3)$ for a triangular face with three spins
$\n_{1,\,2,\,3}$. This should be equivalent to the flux correlations
of the noncompact gauge theory. Indeed the expected long range
correlation functions with the very characteristic dipolar form are
found. Furthermore, this correspondence implies that the ordering
transition of hedgehog-free O(3) model is not in the Heisenberg
universality class, but in the same universality class as the ordering
transition of the NC\cp model. This may be checked by comparing the
universal critical exponents in the two models. Indeed, we find that
the hedgehog-free O(3) model undergoes a continuous ordering
transition with critical exponents that are clearly distinct from the
Heisenberg exponents, but which are consistent with the exponents
obtained in direct Monte Carlo simulation of the transition in the
NC\cp model.  This provides further evidence that the NC\cp model
captures the physics of the hedgehog-free O(3) model.  These exponents
also turn out to be consistent with those obtained from an earlier
attempt to disorder the hedgehog-free O(3) model \cite{KamalMurthy}.

Given the central role played by the NC\cp model, we also present
analytical results on some of the striking properties of this
model. First, we consider the easy-plane deformation of the NC\cp
model when the full spin rotation symmetry is broken down to U(1) by
the presence of easy-plane anisotropy. This model is found to have the
amazing property that under the standard duality transformation, it
maps back onto itself. In particular the zero temperature ordering
transition is found to be self-dual.  Second, we consider the effect
of finite temperature in the NC\cp model. We argue that there is a
thermal Kosterlitz-Thouless phase transition out of the $P^*$
phase. This can be understood as an ionization transition of the
logarithmically interacting spinons. The logarithmic interaction is of
course the Coulomb potential in two spatial dimensions and hence the
presence of such a transition is proof of the existence of gauge
charged particles.

We now briefly comment on the relation of the present paper to earlier
relevant work. Lau and Dasgupta \cite{Dasgupta}, considered the O(3)
model in three Euclidean dimensions on a cubic lattice and applied
complete monopole suppression at every cube of the lattice. This strong
constraint led to the model always being in the ferromagnetic phase,
and the exotic paramagnet was not uncovered in that
work. Subsequently, in an important extension, Kamal and Murthy
\cite{KamalMurthy} allowed for a more flexible definition of
the no monopole constraint by allowing monopole-antimonopole pair
fluctuations if they occurred on neighbouring cubes. In this way, they
were able to obtain a disordered phase, and also found a continuous
ordering transition with non-Heisenberg exponents. However, as pointed
out in that work itself, there are unsatisfactory features of this
algorithm when closely spaced monopoles occur. The problem arises in
the presence of loops of monopole and antimonopoles, where combining
them in pairs is ambiguous. This allows for Monte Carlo moves that
annihilate a monopole and an antimonopole belonging to different pairs,
since the remaining monopoles and anti-monopoles on the loop can be
`re-paired'. This requires making a time consuming nonlocal check for
re-pairing, every time such an event is generated. Consequently,
Kamal and Murthy had to resort to the approximation of making only
local checks of re-pairing. The algorithm then also has the
undesirable property that a given configuration of spins may be
allowed or not allowed depending on the history of how it is generated.
Here we will adopt a more restrictive condition - that monopole
anti-monopole pairs are allowed only if they are {\it isolated} - to
completely circumvent these problems. An improved procedure for
defining the monopole number allows us to easily work with more
complicated lattices. This flexibility will prove very useful in
obtaining the disordered phase. Furthermore, our identification of the
NC\cp model to describe this physics allows us to bring a whole series
of tests to bear on the $P^*$ phase and the transition, which was not
done previously in the absence of such an understanding.

In this paper we will be completely suppressing the free hedgehog
defects, i.e. setting their fugacity to zero. We now briefly discuss
the effect of a finite fugacity, and possible relevance of the physics
described here to systems that may be realized in nature. It is well
known that for pure $U(1)$ gauge theories in D=2+1, introduction of
monopoles leads to confinement \cite{Polyakov}. This result also
implies that the $P^*$ phase is unstable to the introduction of a
nonzero hedgehog (or monopole) fugacity. However, if this fugacity is
small to begin with, deconfinement physics would play an important
role in the finite temperature or short wavelength properties of the
system.

While a finite monopole fugacity necessarily destroys the zero
temperature $P^*$ phase, its effect on the phase transition, where
critical gauge charged bosons are present which act to hinder monopole
tunneling, is less obvious.  In fact, in the sigma model description
of the spin half antiferromagnet on the square lattice, Berry phase
effects \cite{Haldane} lead to a quadrupling of the monopoles
\cite{RS}, which are then more likely to be irrelevant as compared to
single monopole insertions. If monopoles are then irrelevant at the critical
point, the transition could still be controlled by the monopole
suppressed O(3) critical point (or equally the critical NC\cp
model). However, monopole relevance in the adjoining $P^*$ phase
implies that this transition is sandwiched between two conventional
(not deconfined) phases. This dangerously-irrelevant-monopoles
scenario is advocated in a forthcoming paper \cite{Senthiletal}, which
argues that it is possible to have a continuous transition between a
Neel and Valence Bond Solid (VBS) state for the square lattice spin
half antiferromagnet, which is in the same universality class as the
hedgehog free O(3) transition (i.e.~the critical NC\cp theory) studied
here. In fact, Ref. \cite{Senthiletal} argues that the easy plane version of
this transition may already have been seen in the numerical
experiments of \cite{Sandvik}, where there appears to be a continuous
transition between a spin ordered state and a VBS in a square lattice
spin half model with easy plane anisotropy. Such a direct transition
is very natural in the dangerously-irrelevant-monopoles scenario, and
would be controlled by the critical NC\cp model with easy plane
anisotropy, which is also studied in this paper. Thus, the transitions
of the hedgehog suppressed O(3) model, in both the isotropic and easy
plane limits, could be realized in these situations even without
explicit hedgehog suppression, and might potentially be seen directly
in nature. Finally, we note that in D=3+1 a deconfined phase, with
photons and gapped spin half particles, can in principle exist even
with a nonzero hedgehog fugacity.

The layout of this paper is as follows. In Section \ref{supression} we
study the O(3) sigma model with hedgehog suppression. We begin by
describing the particular lattice geometry and hedgehog suppression
scheme used in the Monte Carlo calculations. 
We then present Monte Carlo results that show the presence of a spin
disordered phase in our hedgehog suppressed model.  We argue
that the physics of the hedgehog suppressed sigma model is captured by
the NC\cp model, which implies the presence of photons in this
disordered phase. This leads to the prediction that spin chirality
correlations in the disordered phase take on a very particular long
ranged form which is tested in the Monte Carlo calculations. Next, we
turn to the universal properties of the ordering transition in this
hedgehog suppressed model and find exponents that are distinct from the
Heisenberg exponents. These are then compared with exponents
calculated from directly simulating the NC\cp model. In Section
\ref{variations}, we consider various deformations of the NC\cp model
that correspond to an easy plane anisotropy, a Zeeman field, and the
effect of finite temperature. In particular we prove the remarkable
self duality of the easy plane model. Finally, in Section \ref{3D} we
briefly discuss possible higher dimensional extensions of this
physics.

%%%%%%%%%%%%%%%%%%%%%%%%%%%%%%%%%%%%%%%%%%%%%%%%%%%%%%%%%%%%%%%%%%%%
%%%%%%%%%%%%%%%%%%%%%%%%%%%%%%%%%%%%%%%%%%%%%%%%%%%%%%%%%%%%%%%%%%%%
\section{The Hedgehog Free O(3) Sigma Model}
\label{supression}
\subsection {The Model} 

\begin{figure}
\centerline{\includegraphics[width=3in]{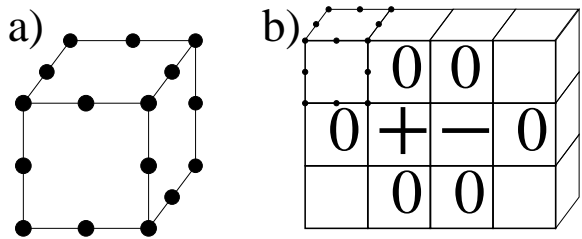}}
\vskip -2mm
\caption{a) Decorated cubic lattice used in the simulations. Spins
live on the lattice points shown, and the monopole number is defined
within each cube. 
b) The only  spin configurations accepted in the simulation are those
that are either hedgehog free, or have hedgehogs that can be 
paired uniquely into isolated nearest neighbor hedgehog-antihedgehog 
pairs. A schematic depiction of such a pairing is shown here in a
vertical section through an isolated pair.}
\label{decmodel}
\end{figure}

We perform Monte Carlo simulations of the three dimensional classical
O(3) sigma model with hedgehog suppression. The lattice that
we consider is a decorated cubic lattice as shown in
Fig.~\ref{decmodel}a, with unit vectors $\n_i$ at the vertices and
edge-centers of the cubic lattice.  As described below, monopole
numbers are associated with the centers of each of these cubes. 
This choice allows for more spin fluctuations when hedgehog
suppression is applied than the simple cubic geometry of
Refs.~\onlinecite{Dasgupta,KamalMurthy}. Neighbouring spins are
coupled via the usual ferromagnetic Heisenberg interaction and hence
allowed states are weighed with the factor $e^{-E}$, with the energy
function $E$ given by:
\begin{equation}
\label{E}
E= -J \sum_{\langle ij \rangle} \n_i \cdot\n_j
\end{equation}

In order to define the monopole number in each cube, we follow reference
\cite{ParkSachdev} and first introduce an auxiliary variable, the
gauge potential $\A_{ij}$ between any pair of neighbouring sites with
spin orientations $\n_i,\,\n_j$. This is defined by introducing an
arbitrary reference vector $\n_*$ and forming the spherical triangle
($\n_*,\,\n_i,\,\n_j$). The edges of a spherical triangle are of
course segments of great circles. If the solid angle subtended by this
spherical triangle is $\Omega[\n_*,\n_i,\n_j]$, then we define:

\begin{eqnarray}
\label{Adef}  
e^{i\A_{ij}} &=& e^{\frac i2 \Omega[\n_*,\n_i,\n_j]} \\ \nonumber
 &=& \frac{1+\n_*\cdot\n_i+\n_*\cdot\n_j+\n_i\cdot\n_j + i\n_*\cdot(\n_i\times\n_j)}{\sqrt{2(1+\n_*\cdot\n_i)(1+\n_*\cdot\n_j)(1+\n_i\n_j)}} ~.
\end{eqnarray}
A different choice of the reference vector, $n'_*$, only leads to a 
`gauge' transformation of $\A$:
$\A_{ij} \to \A_{ij}+\chi_i - \chi_j$,
where $\chi_i = \frac12 \Omega[\n'_*,\n_*,\n_i]$, etc. Thus, gauge
invariant quantities are independent of the choice of the reference
vector. Note also that
$e^{i\A_{ij}}=e^{-i\A_{ji}}$ and these gauge fields are only defined modulo $2\pi$. Thus we have
defined a compact gauge field in terms of the spins. We then define a
flux $F_{\Box}$ on every face bounded by the sites
$(1,2,\dots,n,1)$:
\begin{equation}
e^{i F_{\Box}} = e^{i(\A_{12}+\A_{23}+\cdots +\A_{n1})}
\label{flux}
\end{equation}  
with $F_{\Box} \in (-\pi,\pi]$.  Clearly, the flux is gauge
invariant and hence independent of the choice of reference vector
$\n_*$. The physical meaning of the flux is most readily appreciated by
considering a triangular face with spins $\n_{1,\,2,\,3}$, where it is
approximately the spin chirality:
\begin{equation}
\sin F_{\triangle} \sim  \n_1\cdot(\n_2\times \n_3).
\end{equation}

The hedgehog number $k$ enclosed in a given volume is then the net
flux out of this volume $\sum F_{\Box}=2\pi k$, which is guaranteed
to be an integer from the previous definitions. Note that the hedgehog 
number is simply some function of the spins on a given cube. This
definition is identical to the traditionally used definition of hedgehog
number for volumes that are bounded by triangular faces.  For more
complicated geometries (like the one employed in this work)
however it is a much more natural and powerful definition, 
since it does not rely on an arbitrary triangulation of
the faces and can be quickly computed.

We now consider whether a disordered phase may be obtained while
suppressing the hedgehog configurations. This will favour the
ferromagnetic state which is clearly free of hedgehogs; indeed with
full hedgehog suppression on the simple cubic lattice
\cite{Dasgupta} an ordered phase was found even at zero spin
coupling. In the decorated lattice shown in Fig.~\ref{decmodel}a,
full hedgehog suppression in each cube seems to give rise to a
disordered state for small values of the spin coupling $J$. However,
in order to open a larger window of disordered phase, and obtain more
solid evidence of disorder in the system sizes available, we will
allow for hedgehog-antihedgehog fluctuations on nearest neighbour
(face sharing) cubes. In contrast to Ref.~\onlinecite{KamalMurthy}, we
will only allow for configurations with {\it isolated} hedgehog-antihedgehog
pairs; in other words if a cube contains a hedgehog of strength $q$,
it must contain a nearest neighbour cube with a hedgehog of strength
$-q$ and no hedgehogs in all other nearest neighbour cubes. This is
shown schematically in Fig.~\ref{decmodel}b and gives an unambiguous
prescription for combining the hedgehog and antihedgehogs into
isolated, neutral pairs, and allows us to avoid altogether the
problems in the work of \cite{KamalMurthy}, where such an isolation of
pairs was not demanded.

To summarize, the statistical ensemble is defined as follows.
For each spin configuration, we determine the hedgehog numbers
associated with each cube of the lattice.  If this sample
clears the constraint of no free hedgehogs, 
(we mentioned two versions of this constraint, full suppression 
constraint and the isolated neutral pairs constraint), 
then this configuration is allowed in the ensemble and is weighted
with a relative probability determined by the energy function (\ref{E}).

We simulate this ensemble \cite{fluxsector} using single spin
Metropolis updates in the restricted configuration space.  The data
presented below is taken for 20,000-200,000 Metropolis steps per spin.

%%%%%%%%%%%%%%%%%%%%%%%%%%%%%%%%%%%%%%%%%%%%%%%%%%%%%%%%%%%%%%
\subsection{The Disordered Phase} 
\begin{figure}
\centerline{\includegraphics[width=\columnwidth]{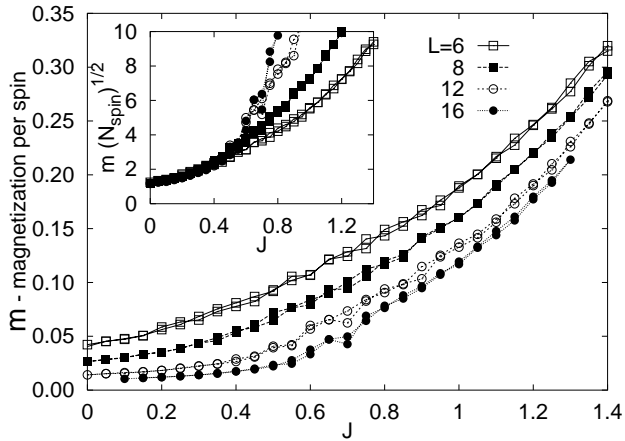}}
\vskip -2mm
\caption{Magnetization per spin, $m = \la |{\bf M}| \ra / N_{\rm spin}$,
with ${\bf M} = \sum_i {\bf n}_i$, as a function of
$J$ for different system sizes 
(we show the data for both sweep directions).
Inset shows the product $m N_{\rm spin}^{1/2}$;
in the magnetically disordered phase, we expect the measured
$\la |{\bf M}| \ra  \sim  N_{\rm spin}^{1/2}$
(for completely uncorrelated spins, the numerical coefficient is 
close to $1$)
}
\label{magn}
\end{figure}

We now discuss the results of the Monte Carlo simulation with hedgehog
suppression. First, in the absence of any hedgehog suppression, the
system is found to have the usual Heisenberg ordering transition at
$J_{\rm c, Heis} \approx 1.7$. Implementing hedgehog suppression that
only allows neutral, isolated pairs of hedgehogs to occur, gives a
smaller but still sizeable region $0 \leq J < 0.7$ over which the system
remains magnetically disordered. This can be seen in Fig.~\ref{magn}
where the magnetization per spin $m$ is plotted for varying system
sizes with linear dimension $L=6,8,12,16$ (the total number of spins
is $N_{\rm spin}=4L^3$).  The magnetization per spin is seen to
approach zero with increasing system size, for small enough values of
$J$. A more convincing demonstration is made in the inset, where we
plot the product of $m$ and the square root of the total number of
spins.  For disordered spins, the average magnetization per spin is
expected to decrease as $N_{\rm spin}^{-1/2}$. Indeed, as seen in the
figure, this situation is realized at least for $J<0.5$.

\begin{figure}
\centerline{\includegraphics[width=\columnwidth]{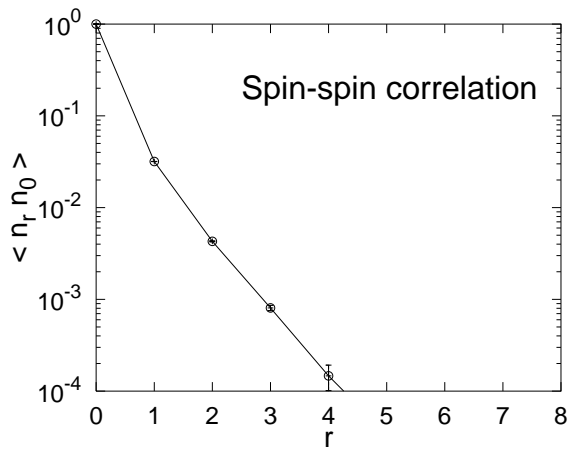}}
\vskip -2mm
\caption{Spin-spin correlations for $J = 0.0$, 
measured for spins at the vertices of the cubic lattice separated by a
distance $r$ along the $\hat z$ direction.  The system size is $L=16$,
so the measurements are done for $r\leq 8$. Note the logarithmic scale
for the vertical axis; the lower cutoff is roughly the limit of what
can be reliably measured in our Monte Carlo.  }
\label{spincorr}
\end{figure}

One may nevertheless worry if there is some other spin order, such as
antiferromagnetic or spiral order, that is not detected by the above
zero-momentum magnetization.  The most direct evidence against any
magnetic order is obtained from the spin-spin correlation, which is
found to be ferromagnetic throughout and rather short-ranged.  For
$J=0$ this is shown in Fig.~\ref{spincorr}, and the spin correlation
indeed decays very quickly, with the correlation length of order one
half lattice spacing.  The above does not mean that the spins are
completely uncorrelated, rather that their correlation is more subtle
as we will see below.

This completes the evidence for the presence of a
magnetically disordered phase $P^*$ with suppressed hedgehogs.
We now investigate the nature of this paramagnetic phase.

%%%%%%%%%%%%%%%%%%%%%%%%%%%%%%%%%%%%%%%%%%%%%%%%%%%%%%%%%%%%%%%%%%
\subsection{Emergent Photons in $P^*$} 

We argue below that the $P^*$ phase of the hedgehog suppressed O(3)
model is distinct from the regular paramagnetic phase $P$ in the model
without such suppression. The sharp distinction arises from the fact
that $P^*$ contains a low energy photon excitation. While this is best
understood by rewriting the O(3) sigma model in the \cp representation,
we first provide a heuristic argument for why such a low energy
excitation may appear before passing to this more complete
explanation.

At any given time slice, the spin configuration (now of spins in a plane)
can be given a skyrmion number. It is easily seen that hedgehog events
change the skyrmion number by unity. Therefore, suppressing hedgehogs
implies that the skyrmion number is a conserved quantity. Thus, if
$j_0$ is the skyrmion density and $j_{1,\, 2}$ are skyrmion currents,
they satisfy the conservation law $\partial_{\mu} j_\mu=0$ with $\mu=0,\,1,\,2$. This
condition may be solved by writing $j_\mu =
\epsilon_{\mu\nu\sigma} \partial_\nu a_\sigma$, in which the skyrmion
current is identified with the flux of a U(1) gauge field. A natural
dynamics would then be given by a Lagrangian ${\mathcal L}=j_\mu
j_\mu$, which would give rise to a linearly dispersing photon.
 
To gain further insight into the nature of the $P^*$ phase, we use
the \cp representation of the O(3) sigma model. It is well known that
the pure O(3) sigma model (no hedgehog suppression) can be rewritten
in terms of a pair of complex bosonic fields $\z= (z_1 \,z_2)^{\rm T}$
that is minimally coupled to a {\it compact} gauge field. The fields
$\z$ transform as spinors under spin rotations, and have unit
magnitude $\z^\dag \z =1$. The spin vector is given by the bilinear
$\n = \z^\dag {\bm \sigma} \z$ (where ${\bm \sigma}$ are the Pauli
matrices), and the flux of the gauge field corresponds to the skyrmion
density of the original spin variables. Compactness of the gauge field
implies the existence of monopoles, which act as sources or sinks of
the gauge flux. These are then to be identified with the hedgehogs
which change the skyrmion number when they occur. Clearly, this \cp
model has two phases, one where the $\z$ particles are `condensed'
which is the ferromagnetic phase (since the gauge neutral unit vectors
$\n$ acquire an expectation value), and another where the $\z$
particles are gapped. The gapped phase is essentially equivalent, at
low energies, to a
pure compact gauge theory which is known to be confining in D=2+1
\cite{Polyakov}. This we associate with the regular paramagnetic
phase.

We now turn to a description of the phases with full hedgehog
suppression within the \cp representation. Indeed, given the
identification of the hedgehog defects with the monopoles of the \cp
theory, hedgehog suppression implies monopole suppression. This is most
directly implemented by passing to a {\it noncompact} gauge field
which is free of monoples. The
distinction between a compact and a noncompact gauge theory cannot be
overemphasized here, since it underlies all the new physics obtained
in this work. The Euclidean action for the noncompact \cp model on a
lattice is given by:
\begin{equation}
S_{{\rm NC}CP^1} = -\frac{\mathcal{J}}{2} \sum_{r,\mu}
     \left( z_r^\dagger z_{r+\hat\mu} e^{i a_{r\mu}} + c.c. \right)
+ \frac{K}{2} \sum_{\Box} ({\bm \Delta} \times {\bm a})^2
\label{noncompact}
\end{equation}
where the lattice curl is the sum of the gauge potentials around a
plaquette. This NC\cp model has two obvious phases - first, a phase 
where the $\z$ particles are `condensed' which is the ferromagnetic 
phase \cite{ferronote}. 
Second, a phase where the $\z$ particles are gapped - this
we identify with the paramagnetic phase $P^*$. However, the noncompact
nature of the gauge field implies that there will be a low energy
photon excitation in this phase. Indeed, within the NC\cp model, the
asymptotic correlation of the flux 
$C_{\mu\nu}= \la F_\mu(r) F_\nu(0) \ra$ in this
phase is simply governed by the free propagation of the photon which
leads to the characteristic dipolar form:
\begin{equation}
C_{\mu\nu}(r) \sim
\frac{3r_\mu r_\nu - \delta_{\mu\nu}r^2}{r^5} ~.
\end{equation}
In particular, for two points separated along the $\hat z$ 
direction, we would expect 
\begin{equation}
C_{zz}(z) \approx \frac{2 B}{z^3} ~,  ~~~~~~~~
C_{yy}(z) \approx -\frac{B}{z^3} ~,
\label{fluxcorr}
\end{equation}
where $B$ is some numerical coefficient.

\begin{figure}
\centerline{\includegraphics[width=\columnwidth]{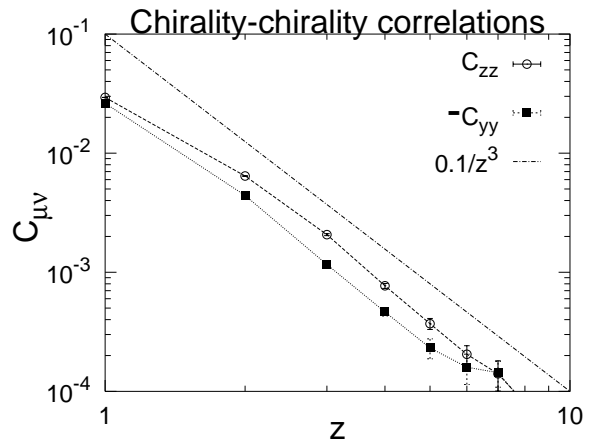}}
\vskip -2mm
\caption{Chirality-chirality (flux) correlations measured along the 
$\hat z$ direction for the same system as in 
Fig.~\ref{spincorr}.  Note the logarithmic scale for both
axes.  We also show a $\sim 1/z^3$ line to indicate the observed 
power law falloff.
}
\label{chiralcorr}
\end{figure}

This prediction of the emergence of a photon in the $P^*$ phase may be
readily checked by using the definition Eq.~(\ref{Adef}, \ref{flux}) of the flux in 
terms of the spins on a face (the spin chirality), and
studying its correlations
\begin{equation}
C_{\mu\nu}(r) \equiv \la \sin F_\mu(r) \sin F_\nu(0) \ra 
\end{equation}
in the hedgehog suppressed sigma model (as usual, $F_{\mu}$ is the flux
through a face perpendicular to $\hat {\bm \mu}$).
The results are shown in Fig.~\ref{chiralcorr} which was taken deep in
the $P^*$ phase with $J=0$.  This figure shows chirality-chirality 
correlations for points separated along the $\hat z$ direction
and corresponds to the prediction in Eqn.~(\ref{fluxcorr}).  
Indeed the expected $1/r^3$ falloff is reproduced, as well as the sign 
of correlations and their approximate relative magnitude.
[For the data in Fig.~\ref{chiralcorr}, we find 
$C_{zz}(z) \approx -1.7\, C_{yy}(z)$; the slight discrepancy is most 
likely because the scaling regime is not quite reached for the 
separation in several lattice spacings.]
Thus, although the spin-spin correlation function is short ranged in 
$P^*$, the spin
chirality correlations have long ranged power law forms. This is a
result of the hidden internal order present in the system that arises
from the suppression of hedgehog defects. It is this internal order
that gives rise to the coherent photon excitation.
(We also note that this topological order survives in a small 
applied Zeeman field $E \to E-h\sum_i n_i^z$, which supports our 
claim for the gapped spinons, --- see Sec.~\ref{sec:zeeman}
below and Ref.~\onlinecite{longpaper}.)

%%%%%%%%%%%%%%%%%%%%%%%%%%%%%%%%%%%%%%%%%%%%%%%%%%%%%%%%%%%%%%%%%%
\subsection{The Transition} 

\begin{figure}
\centerline{\includegraphics[width=\columnwidth]{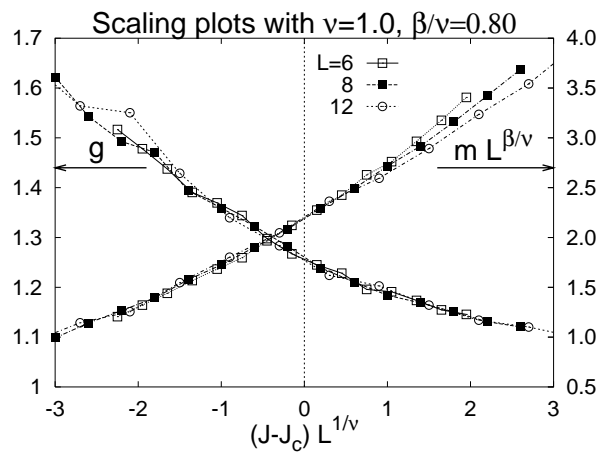}}
\vskip -2mm
\caption{Finite-size scaling plots for the cumulant ratio
(left vertical axis) and magnetization (right axis),
corresponding to the scaling forms~(\ref{gscale})~and~(\ref{mscale}).
We used $J_c=0.725$, $\nu=1.0$, and $\beta/\nu = 0.80$;
the range of the horizontal axis corresponds roughly to 
$J \in [0.40, 1.05]$ for $L=8$ (compare with Fig.~\ref{magn}).
}
\label{scaling}
\end{figure}

We now study details of the ordering transition in the hedgehog
suppressed O(3) model.  We use standard finite size scaling 
analysis in order to estimate the corresponding critical indices.
First, to find the critical point with good accuracy, 
we measure the cumulant ratio (Binder ratio)
\begin{equation}
g = \frac{\la {\bf M}^4 \ra}{ {\la {\bf M}^2 \ra}^2 } ~.
\end{equation}
It is expected to have the finite size scaling form
\begin{equation}
g(J, L) = g(\delta L^{1/\nu}) ~,
\label{gscale}
\end{equation}
where $\delta \equiv J-J_c$ is the deviation from the critical point.
In particular, the curves $g(J,L)$ for different fixed $L$ all 
cross near the critical $J_c$, which we estimate to be 
$J_c = 0.725 \pm 0.025$.  Using the above scaling form, 
we also estimate the correlation length exponent
\begin{equation}
\nu = 1.0 \pm 0.2 ~.  \;\;\;\; (\text{hedgehog suppressed O3})
\end{equation}
The corresponding scaling plot is shown in Fig.~\ref{scaling}.

To estimate the exponent $\beta$, we study the finite size
scaling of the magnetization per spin
\begin{equation}
m(J, L) = L^{-\beta/\nu} f(\delta L^{1/\nu}) ~.
\label{mscale}
\end{equation}
Our best scaling of the data is also shown in Fig.~\ref{scaling},
and we find
\begin{equation}
\beta/\nu = 0.80 \pm 0.05~.  \;\;\;\; (\text{hedgehog suppressed O3})
\end{equation}

We remark that these exponents are consistent with the exponents
obtained in the study by Kamal and Murthy \cite{KamalMurthy} on the
same model but using a different monopole suppression scheme. 
% Kamal and Murthy obtained $\nu = 1.05 \pm 0.05$ and 
% $\beta = 0.75 \pm 0.05$ (thus $\beta/\nu \sim 0.71$).
These exponents are clearly different from the three dimensional 
Heisenberg exponents which are accurately known to be
\cite{ZinnJustin} $\nu_{\rm Heis} = 0.705 \pm 0.003$ and 
$(\beta/\nu)_{\rm Heis} = 0.517 \pm 0.002$.

The specific heat exponent $\alpha=2-d\nu$ is expected to be negative
for this transition ($d$ is the space-time dimensionality), $\alpha
\approx -1.0$, hence a cusp singularity is expected here. Although
such singularities are harder to detect than a divergence, we
nevertheless look for it in the Monte Carlo study of the hedgehog
suppressed model. It is found however that the specific heat remains completely
featureless across the transition; the reason for this behavior is
unclear.

\begin{figure}
\centerline{\includegraphics[width=\columnwidth]{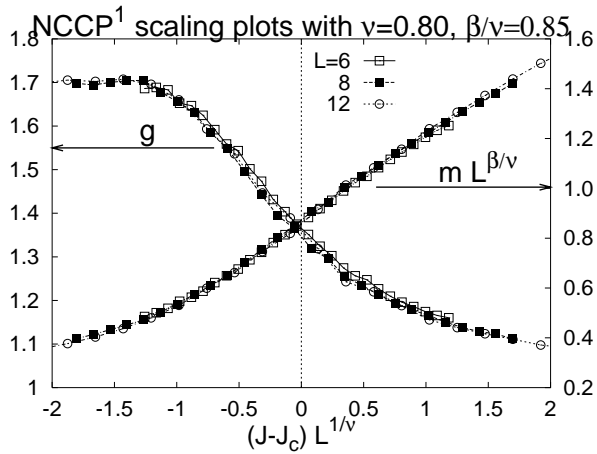}}
\vskip -2mm
\caption{
Finite-size scaling study of the ordering transition
in the NC\cp model Eq.~(\ref{noncompact}).
The system is at $K=0.6$ and the transition from the $P^*$
to the ferromagnetically ordered phase is observed at
$J_c = 1.255 \pm 0.02$; 
the measured order parameter is defined from the field 
$\n_r = \z^\dagger_r {\bm \sigma} \z_r$.
The scaling analysis is similar to Fig.~\ref{scaling}. 
The exhibited plots are for $\nu=0.80$, $\beta/\nu=0.85$;
the horizontal range corresponds to $J\in [1.10, 1.40]$
for the $L=8$ system.
}
\label{nccp1_scaling}
\end{figure}

An important check on the above results in the hedgehog free O(3) sigma
model is provided by comparing with direct simulations on the
noncompact \cp model given by the Euclidean action
(\ref{noncompact}). A complete numerical phase diagram of this model
is reported elsewhere \cite{longpaper}, here we focus on the transition
between the $P^*$ and $F$ phases and comparing the exponents with
those obtained from the hedgehog free O(3) sigma model. Indeed we find
that this transition is second order with exponents
\begin{equation}
\nu=0.8 \pm 0.1 ~, \;\;\;\;
\beta/\nu = 0.85 \pm 0.05 ~, \;\;\;\;  ({\rm NC}CP^1)
\label{expnccp1}
\end{equation}
where the exponent $\beta$ of course describes the onset of ordering
in the gauge neutral field $\n = \z^\dagger {\bm \sigma} \z$. The
corresponding scaling plots for this transition are shown in Figure
\ref{nccp1_scaling}.  Clearly, these are consistent with the exponents
we obtained earlier for the hedgehog free O(3) sigma model, which
leads us to conclude that this transition is indeed distinct from the
Heisenberg ordering transition and is instead described by the
deconfined to Higgs transition in the NC\cp model. The remaining
difference between our best estimates for the critical indices 
[and also between the apparent values of the Binder cumulant at the 
critical points $g(J_c) \approx 1.35$ for the NC\cp transition, and 
$g(J_c) \approx 1.25$ for the hedgehog suppressed O(3) transition]
we attribute to the small system sizes considered. 
For example, for the sizes studied here, the raw data 
(not shown here) Binder cumulant crossing in the NC\cp model has a 
clear downward trend with increasing system sizes, while it changes 
more slowly in the hedgehog suppressed O(3) model; this would narrow 
the difference on going to larger system sizes.

Finally let us note an important physical distinction between this
transition and the Heisenberg transition, which is brought out by
comparing the $\eta$ exponents. This exponent is related to the
anomalous dimension of the $\n$ field and is given by $\eta =
2-d+2(\beta/\nu)$. For the Heisenberg transition, this exponent is
very small $\eta = 0.033 \pm 0.004$. However, for the hedgehog
suppressed O(3) transition we find this exponent to be fairly large
$\eta \approx 0.6$ (and $\eta \approx 0.7$ from direct simulations of
the NC\cp model). A large $\eta$ is to be expected if the magnon field
$\n$ can decay into two deconfined spinons. Indeed, in the limit of
noninteracting spinons, this exponent is expected to approach
unity. Hence, the large $\eta$ obtained at this transition is to be
expected on physical grounds.

%%%%%%%%%%%%%%%%%%%%%%%%%%%%%%%%%%%%%%%%%%%%%%%%%%%%%%%%%%%%%%%%%%%
%%%%%%%%%%%%%%%%%%%%%%%%%%%%%%%%%%%%%%%%%%%%%%%%%%%%%%%%%%%%%%%%%%%  
\section{Deformations of the NC\cp Model. Easy Plane, Zeeman Field and
Effects of Finite Temperature.}
\label{variations}

In this section we consider various deformations of the NC\cp model,
and the effect of finite temperatures. The motivation for this is 
twofold. First, understanding the behaviour of this model in these limits
will provide us with a whole slew of potential checks to further
strengthen the identification between the NC\cp model and the hedgehog
suppressed O(3) model. Some of these explicit checks, like the
correspondence in the easy plane limits and the effect of a Zeeman
field on the isotropic models, will be presented elsewhere
\cite{longpaper}, while others are left for future work. 
Second, we will see below that these models have several interesting
properties and may themselves be directly realizable. For example, the
continuous transition seen in the numerical experiments of
\cite{Sandvik} has been conjectured in \cite{Senthiletal} to be
described by the easy plane NC\cp model.

%%%%%%%%%%%%%%%%%%%%%%%%%%%%%%%%%%%%%%%%%%%%%%%%%%%%%%%%%%%%%%%%%%%
\subsection{Easy Plane Deformation} 
We first consider modeling the easy plane deformation of our O(3)
invariant sigma model with hedgehog suppression. This can be
accomplished, e.g., by having ferromagnetic interaction between
neighboring spins of the form $-J (n_i^x n_j^x + n_i^y
n_j^y)$. Clearly, in this case the spins would prefer to lie in the
$x-y$ plane in spin space; the global spin symmetry of this model is
now broken down from O(3) to $O(2)\times Z_2$, where the $O(2)$
corresponds to spin rotations about the $z$ axis in spin space, and
$Z_2$ arises from symmetry under $n^z\rightarrow -n^z$. The
appropriate deformation of the NC\cp model involves applying the term
$U (|z_1|^4+|z_2|^4)$ at each site with $U >0$ that favours equal
amplitudes for the two components of the spinor field. The
corresponding $\n$ vector will then like to lie in the easy plane. We
will call this the easy plane NC\cp model.

In fact it will be useful to consider the limit of
extreme easy plane anisotropy, where the amplitude fluctuations of the
$\z$ fields are frozen out, and only the phase fluctuations
remain. The $\z$ field can then be parametrized by $z_{1(2)} = \frac
{e^{i\phi_{1(2)}}}{\sqrt{2}} $, and the direction of the spin in the
easy plane is given by $n^x+in^y=2z_1^*z_2=e^{i(\phi_2-\phi_1)}$. The
partition function on a lattice in three Euclidean dimensions can then
be written as:
\begin{eqnarray}
{\mathcal Z}_{EP} &=& \int_{-\infty}^\infty [D {\bf a}]' 
\int_{-\pi}^\pi [D\phi_1 D\phi_2] e^{-S_{EP}} ~, \\ \nonumber
S_{EP} &=& -J \sum_{r, \mu} 
\left[  \cos(\Delta_\mu \phi_1 - a_\mu)
      + \cos(\Delta_\mu \phi_2 - a_\mu) \right] \\ \nonumber
& & + \frac{K}{2} \sum_{\Box} ({\bm \Delta} \times {\bf a})^2 ~,
\label{SEP}
\end{eqnarray}
where the integration over the gauge fields is performed after
suitable gauge fixing to ensure finiteness of the partition function
(hence the prime in the integration measure: $[D {\bf a}]'$).
Below, we will consider applying the standard duality transformations
on this model \cite{DasguptaHalperin,FisherLee}.  For this purpose it is more
convenient to pass to the Villain representation which makes use of
the approximate rewriting $e^{J \cos \alpha} \approx \sum_{n}
e^{in\alpha} e^{-\frac{n^2}{2\mathcal J}}$. Note that the Villain form
has the same $2\pi$ periodicity, $\alpha \rightarrow \alpha + 2\pi$,
as the original partition weight, and for large $J$ we will have
${\mathcal J} \approx J$. The Villain form is expected to retain the
universal properties and phase structure of the original model and
therefore we will start with it and consider a series of exact
transformations that perform the duality. The partition function
written in the dual variables will be seen to be essentially the same
as that written in the direct variables thus establishing the
`self-dual' nature of these models.

The Villain form of the partition function is:
\begin{eqnarray}
{\mathcal Z}_V &=& \int_{-\infty}^\infty [D {\bf a}]' 
\int_{-\pi}^\pi [D\phi_1 D\phi_2]  \sum_{\bj_1,\, \bj_2} 
e^{ -\frac{K}{2} \sum ({\bm \Delta} \times {\bf a})^2 } 
\nonumber \\
& \times &
e^{ -\frac{1}{2{\mathcal J}} \sum (\bj_1^2 + \bj_2^2) }
e^{ i\sum [({\bm \Delta}\phi_1 -{\bf a}) \bj_1
           + ({\bm \Delta}\phi_2 -{\bf a}) \bj_2] } ~,
\label{villain}
\end{eqnarray}
where ${\bf j}_{1,\,2}$ are integer current fields that live on
the links of the lattice.  
We first write ${\mathcal Z}_V$ as a model in terms of the current 
loops only, and then perform an exact rewriting (duality) in terms of 
the vortex loops.  The two forms will have essentially identical
characters when viewed as integer loop models; with proper
identification of the coupling constants, this is the statement 
of self-duality.

We begin by integrating out the phase fields $\phi_{1,\,2}$. This
imposes the condition that the integer current fields 
$\bj_{1,\,2}$ are divergence free. These integer loops are simply 
the world lines of the two bosons $\z_{1,\,2}$, which carry the same 
gauge charge and hence their sum couples to the gauge field. 
Integrating out the gauge field gives rise to a long range Biot-Savart 
type interaction between the total currents. 
The result can be written as:
\begin{eqnarray}
\label{currentloops}
{\mathcal Z}_{V} &=& \sum_{\bj_1,\, \bj_2}
\delta({\bm \Delta} \cdot \bj_1) 
\delta({\bm \Delta} \cdot \bj_2) \nonumber \\
&& \times e^{ -\frac{1}{2} (\bj_1 + \bj_2)_r {\mathcal G}_+(r,r')
                           (\bj_1 + \bj_2)_{r'} } \nonumber \\ 
&& \times e^{ -\frac{1}{2} (\bj_1 - \bj_2)_r {\mathcal G}_-(r,r')
                           (\bj_1 - \bj_2)_{r'} } \\
\label{Gplus}
\tilde{\mathcal G}_+(k) &=& \frac{1}{2{\mathcal J}} 
+ \frac{1}{ 4K(\sin^2 \frac{k_x}2 + \sin^2 \frac{k_y}2
               + \sin^2 \frac{k_z}2) } \\ 
\tilde{\mathcal G}_-(k) &=& \frac{1}{2{\mathcal J}} ~.
\end{eqnarray}
Thus, the original model is equivalent to a model of integer current
loops \cite{Gmunu}, where the combinations $\bj_1 - \bj_2$ have short 
range interactions, while the gauge charged current combinations 
$\bj_1 + \bj_2$ have long ranged interactions with the
asymptotic form 
${\mathcal G}_{+}(R \to \infty) = \frac{1}{4\pi K} \frac{1}{R}$.

We now perform an exact duality transformation on
(\ref{currentloops}). First, the divergence free current loops
can be written as the lattice curl of a vector field
$\bj_{1,\,2} = {\bm \Delta} \times {\bf A}_{1,\,2}/2\pi$ that lives on 
the links of the dual lattice, and the integer constraint is 
implemented using the identity: 
$(2\pi)^3 \sum_{\n} \delta({\bf A} - 2\pi\n)
= \sum_{\bl} e^{i \bl \cdot {\bf A}} $. 
Integrating out the fields ${\bf A}_{1,\,2}$ we obtain the 
following form for the partition function in terms of the dual 
integer current loops $\bl_{1,\,2}$:
\begin{eqnarray}
\label{dualoops}
{\mathcal Z}_{V} &\propto & \sum_{\bl_1,\,\bl_2}
\delta({\bm \Delta} \cdot \bl_1)
\delta({\bm \Delta} \cdot \bl_2)  \nonumber \\
&& \times e^{ -\frac{1}{2}(\bl_1 + \bl_2)_r {\mathcal F}_+(r,r')
                          (\bl_1 + \bl_2)_{r'} }  \nonumber \\
&& \times e^{ -\frac{1}{2}(\bl_1 - \bl_2)_r {\mathcal F}_-(r,r')
                          (\bl_1 - \bl_2)_{r'} }  \\
\tilde{\mathcal F}_+(k) &=& 
\frac{\pi^2 K {\mathcal J}}
     {{\mathcal J} + 2K (\sin^2 \frac{k_x}2 + \sin^2 \frac{k_y}2
                         + \sin^2 \frac{k_z}2)}  \\
\tilde{\mathcal F}_-(k) &=& 
\frac{\pi^2 {\mathcal J}}
     {2 (\sin^2 \frac{k_x}2  + \sin^2 \frac{k_y}2 
         + \sin^2 \frac{k_z}2)} ~.
\end{eqnarray}
One can argue that the two integer loops of this dual representation
(\ref{dualoops}) are precisely interpreted as the vorticities in the
original two boson fields.  A vortex of equal strength in both boson
fields ($\bl_1 = \bl_2$) can be screened by the gauge field, and hence
has only short range interactions, while unbalanced vortices which
cannot be completely screened interact via long range
interactions. Correspondingly, the combination of the dual currents
$\bl_1 + \bl_2$ interact via a short range interaction ${\mathcal
F}_+(R)$, while the other combination $\bl_1 - \bl_2$ has a long
range interaction with the same asymptotic form as in (\ref{Gplus}),
${\mathcal F}_{-}(R\to \infty) = \frac{\pi\mathcal J}{2}
\frac{1}{R}$. Indeed the partition function written in terms of dual
loops (\ref{dualoops}) is essentially the same as when written in
terms of the direct variables (\ref{currentloops}), which can be seen
by making the association $\bl_1 \leftrightarrow \bj_1$ and $\bl_2
\leftrightarrow -\bj_2$. The only differences are in the form of the short range interactions which do
not affect universal properties.  As argued below this immediately
implies that the transition between $P^*$ and $F$ phases in this model
will be self-dual.  Of course, we can also rewrite (\ref{dualoops}) in
terms of the fields conjugate to the currents $\bl_{1,\,2}$, obtaining
the action in terms of the dual (vortex) fields and the dual gauge
field, which will have essentially the same form as the original action 
Eq.~(\ref{villain}).  We do not spell this out here since the exhibited 
forms already suffice.

We first note the description of the various phases in terms of the
properties of boson current loops and also the vortex (dual) loops. In
the direct representation, when the partition function is dominated by
small loops of $\bj_1 \pm \bj_2$, the system is in the `insulating' or
paramagnetic state $P^*$. In this phase there is a single low energy
excitation - the photon. On the other hand, when these loops become
arbitrarily large, which certainly occurs if both ${\mathcal J},\,K
\gg 1$, we are in the `superfluid' or ferromagnetic phase $F$. Here
too we have a single low lying mode, the magnon, which is the
Goldstone mode arising from the spontaneously broken spin symmetry
within the easy plane.

The two integer loops of the dual representation (\ref{dualoops}) 
correspond to the world lines of the vortices in the
two boson fields. It is easily seen, for example by analyzing the dual
action in different parameter extremes, that the $P^*$ phase occurs if
{\it large} vortex loops of both $\bl_1 \pm \bl_2$ proliferate,
while if both the vortex loops are typically small, the $F$ phase
results. 

In terms of the direct boson variables, the $F$ phase is the ordered
phase, and hence has a low energy Goldstone mode, while the $P^*$
phase is the disordered phase and has a low energy photon mode. In the
dual variables the roles are reversed - the $F$ phase is the disordered
phase with the dual photon, while the $P^*$ phase is the dual
`ordered' phase, with the Goldstone mode.  In particular, a direct
transition between $P^*$ and $F$ can either be thought of as an
ordering transition in the direct variables, or the reverse in terms
of the dual variables. This interchange of ordered and disordered
phases is typical of a duality transformation. What is special in this
case is that the partition function is essentially identical when
written in terms of the direct and dual variables. This implies that
the transition must be self-dual. One direct consequence of this self
duality is that the critical exponent ($\beta$) measuring the rise of
the order parameter on the ordered side of the transition is equal to
that of a suitably defined {\it disorder} parameter on the opposite
side of the transition. Further consequences arising from the 
self duality and other phases contained in this easy plane NC\cp model
will be discussed at length elsewhere \cite{longpaper}, along with
results of Monte Carlo studies.  Here, we just note that a continuous
transition from the $P^*$ to $F$ phase is obtained, with critical
indices: 
\begin{equation}
\nu=0.60 \pm 0.05 ~, \;\;\;
\beta/\nu = 0.70 \pm 0.05 \;\;\;  (\text{easy-plane}).
\label{easyplaneexp}
\end{equation}

Finally, we note that this critical point is conjectured in 
\cite{Senthiletal} to control the continuous transition seen in the
numerical experiments of \cite{Sandvik}.

%%%%%%%%%%%%%%%%%%%%%%%%%%%%%%%%%%%%%%%%%%%%%%%%%%%%%%%%%%%%%%%%%%%%%
\subsection{Effect of Zeeman Field}
\label{sec:zeeman}
We now consider applying a uniform Zeeman field along the $n^z$
direction in spin space, $S_{Z}= -h\sum_i n^z_i$, and study the effect
on the various phases and phase transitions in both the isotropic and
easy plane models. In the \cp representation this term can be written
as:
\begin{equation}
S_Z = -h\sum (|z_1|^2-|z_2|^2)
\label{Zeeman}
\end{equation}
which breaks the $z_1 \leftrightarrow z_2$ and the only remaining
symmetry is that of $U(1)$ spin rotations about the $n^z$ axis.

\begin{figure}
\centerline{\includegraphics[width=3.6in]{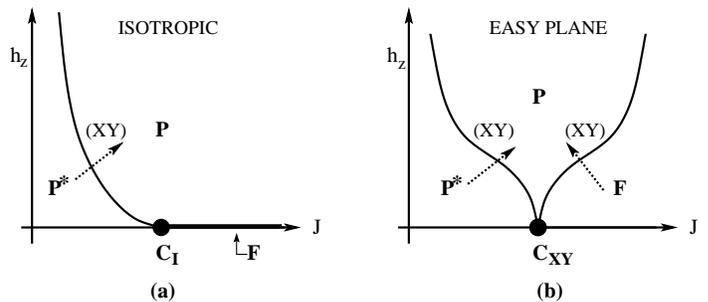}}
\vskip -2mm
\caption{Effect of a Zeeman field on the zero temperature phases of
the NC\cp model. (a) Isotropic case - Two distinct paramagnetic phases
exist in the presence of a Zeeman field. The $P^*$ phase has a low
lying excitation, the photon, while the $P$ phase is gapped and breaks
no global symmetries. The two are separated by a continuous transition
in the 3D XY universality class (the arrow indicates the ordered to
disordered 3D XY transition). At zero field the transition $C_I$ is
described by the critical NC\cp model. (b) Easy Plane case - Zeeman
field is applied perpendicular to the easy plane. Spins can order in the
easy plane to give the $F$ phase which has a low lying excitation, the
Goldstone mode. This is separated from the disordered $P$ phase by a
3D XY transition. There are two paramagnetic phases, $P^*$ with a
photon and a gapped $P$ phase separated by a 3D XY transition. At zero
Zeeman field the two 3D XY transitions meet to give the self dual
$C_{XY}$ transition.}
\label{zeemanfig}
\end{figure}

For the model with the isotropic exchange coupling, adding the term
(\ref{Zeeman}) to the action (\ref{noncompact}) will result in the
phase diagram shown in Figure (\ref{zeemanfig}a). At zero field we
have the $P^*$ and $F$ phases separated by the monopole suppressed
O(3) transition, or equivalently the NC\cp ordering transition
$C_I$. Turning up the Zeeman field, the ordered moment of $F$ locks
into the field direction - the resulting phase does not break any of
the global symmetries and has no low lying modes. Hence we will call
this phase $P$. (Note that for the isotropic NC\cp model at zero
temperature, this phase is only present when a Zeeman field is
applied). This phase can be understood as a Higgs phase with
condensation of the $z_1$ field but not of the $z_2$ field.

On the other hand, in the $P^*$ phase there is a gap to
$z_1$ and $z_2$ particles. Turning up the Zeeman coupling lowers the
$z_1$ branch until at the transition it condenses to give the $P$
phase. Thus, the transition between the $P^*$ and $P$ phases involves
condensing a single scalar field coupled to a gauge field and hence is
expected to be in the inverted-XY universality class
\cite{DasguptaHalperin} (i.e. the $P^*$ to $P$ transition is the
ordered phase to disordered phase transition of the 3D XY model). 

The shape of the phase boundary near the zero field transition can be
related to the critical exponents of the NC\cp model. In particular it
depends on the ratio of the scaling dimension of the Zeeman operator
($y_h=d-\frac{\beta}{\nu}$ for the isotropic case) to the scaling dimension of the `thermal'
operator ($y_t =\frac1\nu$). Thus, if $\delta J$ is the deviation away
from the zero field critical point $C_{I}$, then the phase boundary
is given by the curve 

\begin{eqnarray}
h_z &\propto& (\delta J)^\phi \\
\phi &=& y_h/y_t 
\label{curve}
\end{eqnarray}
. The expressions for the scaling dimensions yield $\phi=y_h/y_t =\nu
d - \beta$. Using the numerical values of the exponents in
Eq.~(\ref{expnccp1}), we have $\phi \approx 1.7$, which gives a phase
boundary that approaches the isotropic transition point horizontally
as shown in Figure \ref{zeemanfig}a.
\vskip 2mm

For the model with easy plane exchange coupling, the effect of adding
the Zeeman field perpendicular to the easy plane can be argued to lead
to the phase diagram in Figure \ref{zeemanfig}b. Here we avoid using
the extreme easy plane limit, where $n^z$ was set identically to zero,
in order to have a finite coupling to the applied
field. (Alternatively, we can model the effect of the Zeeman field by
considering an action similar to Eq.~\ref{SEP} but with different
couplings $J_1$ and $J_2$ for the two angle variables.)  At zero field
we have the $P^*$ and $F$ phases separated by the monopole suppressed
easy plane transition.  Again, in the presence of a Zeeman field, a
$P$ phase is possible, with no spontaneously broken symmetry and no
low lying excitations, where only the $z_1$ field is condensed. The
phase transition between $P^*$ and $P$ is then driven by the
condensation of the gauge charged $z_1$ field, and is hence expected
to be in the same universality class as the ordered to disordered
transition of the 3D XY model \cite{DasguptaHalperin}. Starting in the
$P$ phase, as the exchange coupling is further increased, it becomes
favourable for the $z_2$ field to also condense, so that there develops a 
finite expectation value for the in-plane spin operators $n^x+in^y =
2z_1^* z_2$. This leads to the $F$ phase, which spontaneously breaks the
remaining spin rotation symmetry and has one Goldstone mode. Since the
gauge field has already been `Higgsed' away, this transition is the
regular ordering transition of the 3D XY model. As described
previously, the original easy plane model is essentially self dual,
and this remains true on adding the Zeeman term. This explains the
`reflection' symmetry of the universal properties in the phase diagram
- for example the duality interchanges the $P^*$ and $F$ phases while
the $P$ phase is taken to itself. This also implies that the
transition from $P^*$ to $P$ must be in the same universality class as
the transition from $F$ to $P$, which was indeed the result of the
previous analysis, which finds them both to be in the 3D XY
universality class. 

The shape of the phase boundaries near the zero field easy plane transition can
be related to the critical exponents of the easy plane NC\cp model
using Equation \ref{curve}. While the thermal eigenvalue in this case
is already known (Eqn. \ref{easyplaneexp}), the scaling dimension of
the Zeeman operator needs to be determined. This is conveniently done
within the easy plane NC\cp model at criticality, by studying the
scaling dimension of the operator that gives rise to unequal hopping
strengths $J_1 \neq J_2$ for the two species of bosons in
Eqn. \ref{SEP}. To this end, consider defining the link operators:
\begin{equation}
O^{\pm}_{ij}(r) = \{ \cos (\Delta \phi_1 - a) \pm \cos (\Delta \phi_2 -a)\}
\end{equation}
Clearly, adding $\sum_{\langle ij \rangle} O^-_{ij}$ will give rise to
unequal hoppings, and hence has an overlap with the Zeeman
operator. Similarly, $\sum_{\langle ij \rangle} O^+_{ij}$ will have an
overlap with the `thermal' operator. Thus, studying the power law
decay of the correlators of $O^{\pm}_{ij}$ at criticality allows us to
extract the scaling dimensions of the Zeeman and thermal
operators. The latter may be compared against other, more accurate
determinations of the same quantity and serves as a check of this
approach. Thus we find:
\begin{eqnarray}
y_h &=& 1.2 \pm 0.3\\
y_t &=& 1.6 \pm 0.2
\end{eqnarray}
the value of $y_t$ obtained is consistent with the value of $\nu$ quoted in
Eqn. \ref{easyplaneexp} from which $y_t \approx 1.7 $. Thus we have
$\phi = 0.75 \pm 0.2$. Note that since $0<\phi<1$, the shapes of the
phase boundaries are as shown in Figure \ref{zeemanfig}b \cite{leon}.

The NC\cp representation of the hedgehog-free easy-plane model in 
the presence of a Zeeman field also arises in the context of two component superconductors,  as described in Ref. \cite{Babaev} where the continuum version of this identification was also noted.

Finally we note that these properties of the easy plane model in the
presence of a perpendicular Zeeman field are potentially useful to
testing the conjecture of \cite{Senthiletal} that the continuous
transition seen in the numerical experiments of \cite{Sandvik} is
controlled by the critical easy plane NC\cp model. An important point
is that given the mapping of the Neel field to the unit vector of the
sigma model, the Zeeman field considered here actually corresponds to
a {\it staggered} Zeeman field on the spins of Ref.~\onlinecite{Sandvik}.

%%%%%%%%%%%%%%%%%%%%%%%%%%%%%%%%%%%%%%%%%%%%%%%%%%%%%%%%%%%%%%%%%
\subsection{Finite Temperatures} 
We now investigate the finite temperature properties of the NC\cp
model (or equivalently, the monopole suppressed O(3) sigma model),
both for the isotropic as well as the easy plane case. The main result
is that there is a finite temperature version of the $P^*$ phase,
which we call the thermal Coulomb phase $P_T^*$, with power law
correlations of the electric fields. This is distinct from the regular
paramagnetic phase $P$ with short ranged correlations, and is
separated from it by a finite temperature Kosterlitz-Thouless phase
transition. The existence of such a transition can be seen from the
following physical argument. In the zero temperature $P^*$ phase,
there are gapped spinons that interact via a logarithmic
interaction. If this interaction persists to finite temperatures, then
clearly for small enough temperatures, thermal fluctuations will only
generate neutral spinon pairs. This is the $P^*_T$ phase. However, at
some higher temperature it becomes entropically favourable to
proliferate unbound spinons (exactly as with logarithmically
interacting vortices in the two dimensional XY model) and a
Kosterlitz-Thouless type transition to a spinon-plasma phase will be
expected. Indeed, for the hedgehog suppressed O(3) model, the
existence of such a finite temperature transition may be viewed as
evidence for the existence of emergent gauge charged objects
interacting via a Coulomb potential, that happens to be logarithmic in
two spatial dimensions which gives rise to the transition.

\begin{figure}
\centerline{\includegraphics[width=3.6in]{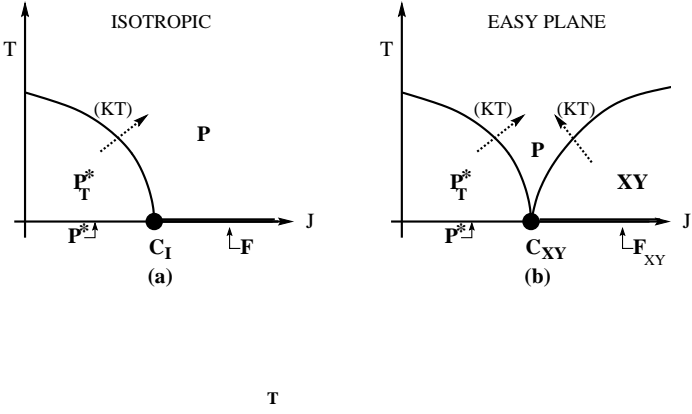}}
\vskip -2mm
\caption{Effect of finite temperature $T$ on the NC\cp model. (a)
Isotropic case:  The thermal Coulomb phase $P^*_T$,  a finite
temperature analogue of $P^*$, has power law electric field
correlations (Eq. \ref{efield}) and is separated from
the more conventional $P$ phase by a Kosterlitz-Thouless phase
transition.  This can be understood as the thermal ionization of
spinons interacting via the 2D Coulomb (log) potential. The $F$ phase
is disordered at any finite temperature. (b) Easy Plane
case:  The ordered phase now has algebraic correlations at finite
temperature ($XY$) which is also separated from the $P$ phase by a 
KT transition.
}
\label{finiteT}
\end{figure}

We now sketch a derivation of these results - we choose to do this in
a model with a single species of `spinon' or gauge charged
particle.  Treating this single component (or noncompact $CP^0$) model
will simplify the discussion and the very same results hold for the
models of interest, i.e. the isotropic and easy-plane NC\cp
models, because it is only the logarithmic binding/unbinding of 
the gauge charged objects that matters.
Since the finite temperature physics described above occurs in
the charge and electric field sector, it will be useful to consider
writing the thermal partition function $\mathcal{Z}_\beta =
{\rm Tr} \,e^{-\beta H}$ in terms of the zero Matsubara
frequency components of these fields.  
The appropriate effective energy is:
\begin{equation}
E\{n_r ,e_{r\mu}\} = \frac{U}{2}\sum_r n_r^2 + 
\frac{1}{2\mathcal{K}} \sum_{r\mu} e_{r\mu}^2
\end{equation}
where $r$ and $\mu$ run over the sites and directions of the two 
dimensional spatial lattice.
The partition function involves a sum over all configurations of
the integer charge fields $n_r$ and {\it real} electric fields 
$e_{r\mu}$ (since we are working with a noncompact gauge theory) 
that satisfy the Gauss law constraint 
${\bm \Delta} \cdot {\bf e}= n_r$ at each site of the lattice.
Note that the boson number and electric field variables that we
use here are conjugate to the phase and gauge potential variables 
that have been used so far (cf.~Eqs.~\ref{noncompact},\ref{SEP}).
Our treatment here is precise in the regime where the total 
Hamiltonian is dominated by the exhibited $E\{n_r, e_{r\mu}\}$ part,
but is also valid more generally in the effective sense.

The Gauss law constraint can be implemented 
using a Lagrange multiplier $a_0$ to give the following expression for 
the partition function:
\begin{equation}
\mathcal{Z}_\beta \sim 
\sum_{\{n_r\}} \int [D{\bf e}] \, \int [Da_0] 
e^{- \beta E\{n_r, e_{r\mu}\} + i \sum a_0({\bm \Delta}\cdot {\bf e}-n)}
~.
\end{equation}
To establish that there is a thermal transition, we integrate out the
fields $a_0(r),\, e_{r\mu}$, so the partition weight is now just a
function of the integer charges. This yields:
\begin{equation}
\mathcal{Z}_\beta  \sim  \sum_{\{n_r\}} 
e^{- \frac{\beta}{2} \sum_{rr'} n_r V_{rr'} n_{r'}}
\end{equation}
where for large separation the potential between charges takes the form
$V_{rr'} \sim \frac{1}{2\pi{\mathcal K}} \log \frac{1}{|r-r'|}$, which
is the Coulomb interaction in two dimensions. It is well known that
such a Coulomb gas has two phases, one with the charges bound into neutral
pairs at low temperatures, and a plasma phase at high temperatures
separated by a KT transition which occurs at $T_c \le (4 \pi{\mathcal
K})^{-1}$.  This is not surprising since we know that NC$CP^0$
model is dual to the O(2) quantum rotor model in two spatial
dimensions, and the spinons of the NC$CP^0$ map onto the vortices of
the O(2) rotor model, which interact logarithmically.

A sharp distinction between the low and high temperature paramagnetic
phases can be drawn by looking at the electric field correlators. In
the low (but nonzero) temperature $P^*_T$ phase, these fall off as the
inverse square of the distance, while in the high temperature `spinon
plasma' phase $P$, these correlators are short ranged:
\begin{equation}
\la e_\mu(r) e_\nu(0) \ra \sim  
\Bigl\{ 
\begin{matrix} 
\frac{1}{r^2} (\frac{2r_\mu r_\nu}{r^2} - \delta_{\mu\nu}) & \;T<T_c \\ 
\text{short ranged} & \;T>T_c
\end{matrix}
\label{efield}
\end{equation}
where the indices $\mu,\, \nu \in\{x,y\}$. As shown elsewhere
\cite{longpaper}, exactly the same results are obtained for the
isotropic NC\cp model, and the phase diagram is shown in 
Fig.~\ref{finiteT}a. In particular, despite the presence of a global 
SU(2) symmetry, the transition remains KT. 

For the easy plane NC\cp model, the expected finite temperature phase
diagram is shown in Fig.~\ref{finiteT}b.  Here, besides the discussed
$P^*_T$ and $P$ phases, there is also a power law phase ($XY$) that
appears out of the zero-temperature ferromagnetically ordered phase
and terminates at the usual KT transition to the $P$ phase. Again, the
similarity between the $P^*_T$ and $XY$ sides of this phase diagram have
to do with the self duality of the underlying easy-plane NC\cp model.

%%%%%%%%%%%%%%%%%%%%%%%%%%%%%%%%%%%%%%%%%%%%%%%%%%%%%%%%%%%%%%%%%%
%%%%%%%%%%%%%%%%%%%%%%%%%%%%%%%%%%%%%%%%%%%%%%%%%%%%%%%%%%%%%%%%%%
\section{Towards D=3+1 and Layered Phases}
\label{3D}
All of the preceding discussion was focused on the D=2+1 dimensional
system where complete hedgehog suppression was required in order to
obtain the deconfined phase. It is natural to ask whether similar
physics can be obtained in models with only a finite energy cost for
hedgehogs. This can occur in the presence of critical bosonic fields
as argued in \cite{Senthiletal} leading to deconfined quantum criticality
in D=2+1 dimensional systems. Another way in which a deconfined phase
can be stabilized with a finite hedgehog fugacity is to consider D=3+1
dimensional systems. Now, the hedgehog is a particle, and a finite
hedgehog fugacity does not necessarily destabilize the Coulomb
phase. Thus, in principle one could obtain a D=3+1 U(1) deconfined
phase with global SU(2) symmetry, by disordering the O(3) sigma model
with a sufficiently large but finite hedgehog core energy. This
question is currently under investigation; the obvious extensions of 
the presented D=2+1 dimensional realization either led to ferromagnetic 
order or to the conventional paramagnet.

An interesting possibility that we remark upon is a layered system
that lives in three spatial dimensions, where each layer realizes the
D=2+1 dimensional deconfined phase. Such a phase exhibits {\it quantum
confinement} in that the gauge charged spinons can move freely within
a layer but not between the layers, and the photon remains a D=2+1
dimensional particle. This is reminiscent of the sliding Luttinger
liquid phase \cite{SLL} of coupled one dimensional systems that also
exhibits the phenomenon of quantum confinement, and has a low energy
phonon excitation (which is the analogue of the photon in this lower
dimensional system).

%%%%%%%%%%%%%%%%%%%%%%%%%%%%%%%%%%%%%%%%%%%%%%%%%%%%%%%%%%%%%%%%%%%%
%%%%%%%%%%%%%%%%%%%%%%%%%%%%%%%%%%%%%%%%%%%%%%%%%%%%%%%%%%%%%%%%%%%%
\section{Conclusions}
We conclude by highlighting the main results of this work.  We
considered the question of whether the O(3) sigma model can be
disordered without monopoles, and answered it in affirmative.  We
provided an explicit example of an O(3) Heisenberg spin system with no
free hedgehogs and no magnetic order in (2+1)D.  
While this is in agreement with the earlier work of Kamal and 
Murthy\cite{KamalMurthy}, the potentially problematic features of the 
procedure adopted in that work have been entirely avoided in the 
present paper. 

Furthermore, we identified the proper description of the hedgehog
suppressed O(3) model which involves spinons coupled to a noncompact
gauge field, which may be called the {\it noncompact} \cp model. Thus
the hidden topological order resulting from hedgehog suppression gives
rise to a photon-like low energy excitation in the paramagnetic 
($P^*$) phase of this model, which leads to power law correlations of
the spin chirality.  This may also be viewed as an example of a U(1) 
fractionalized phase (albeit with complete monopole suppression)
with full SU(2) spin rotation symmetry.

Our understanding of the hedgehog suppressed disordered sigma model is
interesting from the statistical mechanics point of view and addresses
some long standing questions.  In a sense, we identified how to
``decompose'' the O(3) model into a part that involves the topological
defects (hedgehogs), and the part which does not involve these.  Such
decomposition of the O(2) model into the vortex and spin wave parts is
well known.  The corresponding ``spin wave'' part for the O(3) model
turns out to be the NC\cp model, which clearly has little to do with
spin waves.  In particular, the commonly asked question whether the
spin waves (perhaps non-linearly coupled) can disorder the O(3) model
seems to have no meaning.

Another question has been on the role played by the hedgehog
defects at the Heisenberg transition. A sharp formulation of this
question is whether the ordering transition in the hedgehog suppressed
model is identical to the Heisenberg transition. Our calculations of
the universal critical exponents for this transition ($\nu=1.0\pm 0.2$
and $\beta/\nu = 0.80 \pm 0.05$) show that it is indeed distinct from
the Heisenberg transition. Direct simulation of the NC\cp model
however yields exponents that are consistent with these. Moreover, the
large $\eta \approx 0.6$ of the vector (magnon) field implied by these
exponents can be heuristically understood since the magnon can decay
into a pair of unconfined spinons at this critical point.

Thus, an important conceptual issue that is clarified by our work is
the existence of two different spin rotation symmetric critical points
in D=2+1. The first is the Heisenberg transition whose `soft spin'
field theory is given by the O(3) $\phi^4$ theory. This describes of
course the transition in the O(3) sigma model if it is regularized by
putting the spins on the lattice. It also describes the transition in
the lattice \cp model with a {\it compact} gauge field. The second
transition is described by a `soft spin' field theory with a pair of
complex scalar fields (that transform as spinors under spin
rotations), coupled to a {\it noncompact} gauge field. This describes
the ordering transition of the lattice O(3) sigma model with hedgehog
suppression. It also describes the transition in the lattice \cp
coupled to a {\it noncompact} gauge field.  Earlier indiscriminate use
of what is effectively the NC\cp model to describe the Heisenberg
transition are therefore to be reconsidered.

\vskip 2mm
We also studied various physical extensions of the hedgehog suppressed
O(3) model, such as adding Zeeman field and also the effect of finite
temperature.  The $P^*$ phase survives in small external magnetic
field and retains the photon, which is an evidence for gapped spinons.
At finite temperatures, a thermal Coulomb phase with long ranged
power law electric field correlations exists, and undergoes a
Kosterlitz-Thouless transition to the usual paramagnetic phase at
higher temperatures; this is indirect evidence for gauge charge
carrying particles (spinons) that interact via the Coulomb potential 
which is logarithmic in 2D.

An interesting extension is obtained by breaking the full rotation
symmetry down to the easy-plane XY symmetry.  We showed that this
model possesses a remarkable `self-duality' property. Indeed as far as
we know this is the first purely bosonic model in D=2+1 that displays
this property. In particular the critical point describing the
ordering transition in this model is self-dual. An important role is
played by this critical theory in a forthcoming publication
\cite{Senthiletal}, where it is conjectured that this critical point
may already have been seen in numerical experiments on easy plane spin
half quantum antiferromagnet on the square lattice where surprisingly
a continuous transition is observed between a spin ordered state and a
valence bond solid that breaks lattice symmetries \cite{Sandvik}.

\vskip 2mm
In D=2+1 the $P^*$ phase and its gapless photon excitation only exist
in the limit of infinite hedgehog suppression. Thus, within this phase,
the complete suppression of free hedgehogs represents an extreme limit
that may seem unnatural. However, reasoning in this limit can be
conceptually powerful and throw light on many tricky issues. Besides, we
note that even if hedgehogs eventually proliferate and lead to a gap
for the photon, this $P^*$ phase may be relevant for describing
physical crossovers in some strongly correlated systems at energy
scales above the photon mass. In contrast to the $P^*$ phase which in
the \cp language is unstable to turning on a finite monopole fugacity,
the stability of the critical points where gapless gauge charged
particles are present is a more involved question, and depends on the
number of gapless fields present and the particular monopole creation
operator under consideration. In fact, it is argued in
\cite{Senthiletal} that quadrupled monopole operators (which are
relevant to the spin half quantum antiferromagnets on the square
lattice) are irrelevant at the NC\cp critical point in both the
isotropic and easy plane limits. This would allow for a continuous
transitions between valence bond solid and Neel phases in such
systems, and these transitions would be controlled by the critical
points studied in this paper. Finally, we expect a similar $P^*$ phase
in the (3+1)D O(3) model, which will now be present also with strong
but finite monopole suppression, which would provide a spin rotation
invariant (3+1)D model which exhibits a fractionalized Coulomb phase
with deconfined spinons and a true photon excitation.

%%%%%%%%%%%%%%%%%%%%%%%%%%%%%%%%%%%%%%%%%%%%%%%%%%%%%%%%%%%%%%%%%%
\section{Acknowledgements}
We thank L. Balents, T. Senthil, S. Minwalla and S. Sachdev for
stimulating discussions, and D. A. Huse for useful suggestions and for
generously sharing with us his unpublished work. We are especially
grateful to M. P. A. Fisher for several vital conversations at
beginning of this project and for his encouragement throughout. OIM
was supported by the National Science Foundation under grants
DMR-0213282 and DMR--0201069. AV would like to acknowledge support
from a Pappalardo Fellowship.

%%%%%%%%%%%%%%%%%%%%%%%%%%%%%%%%%%%%%%%%%%%%%%%%%%%%%%%%%%%%%%%%%%


\begin{thebibliography}{x}
\bibitem{nccp1} A more accurate name would be the
two-component SU(2) symmetric lattice superconductor
model. Nevertheless we proceed with the term NC\cp, since it provides
a short and convenient mnemonic for the matter content of the model
and the crucial noncompact nature of the gauge field.
\bibitem{fluxsector} In a system with monopole suppression
and periodic boundary conditions as used in our Monte Carlo, local
spin updates do not change the total flux in each direction. Our
simulations occur in the zero flux sector since we start with the
fully polarised state. Averages taken in this sector and averages
taken over all sectors are expected to equal each other in the
thermodynamic limit.
\bibitem{ferronote} Note that the excitation spectrum of this
ferromagnetic phase in the noncompact model differs in
subtle ways from that of the ferromagnetic phase in the compact \cp
model, but the ground state properties which we focus on here, are
identical.
\bibitem{Gmunu}
Note that the loop interaction ${\mathcal G}_{r\mu, r'\mu'}$
is defined only up to 
${\mathcal G}_{r\mu, r'\mu'} \to {\mathcal G}_{r\mu, r'\mu'}
+ \partial_{r_\mu} \alpha(r,r') + \partial_{r'_{\mu'}} \beta(r,r')$
with arbitrary $\alpha(r,r')$ and $\beta(r,r')$, since this leaves 
the interaction energy unchanged because of the divergence free 
condition.
\end{thebibliography}
\end{document}